\documentstyle[12pt,aaspp]{article}
\def\h{$h^{-1}$}
\def\Mpch{h^{-1}{\rm Mpc}}
\def\Mpc{{\rm Mpc}}
\def\etal{et al. }
\lefthead{Walter \& Klypin}
\righthead{Evolution of clusters}

\begin{document}

\title{Evolution of Clusters in Cold plus Hot Dark Matter Models}
\author{Charles Walter and Anatoly Klypin}
\affil{Department of Astronomy\\ New Mexico State University \\
Las Cruces, NM 88003-0001}

\begin{abstract}
We use N-body simulations to study evolution of galaxy
clusters over the redshift interval 0 $\le$ z $\le$ 0.5 in
cosmological models with a mixture of cold and hot dark matter (CHDM).
Four different techniques are utilized: the cluster-cluster
correlation function, axial ratios and quadrupoles of the dark matter
distribution in individual
clusters, and virial properties. We find that the correlation function
for clusters of the same mass limit was larger and steeper at high
redshifts.  The slope  increases from 1.8 at $z=0$ to 2.1 at $z=0.5$.
Comoving correlation length $r_c$  scales with the mass limit $M$
within comoving radius $1.5\Mpch$ and the redshift $z$ as
$r_c\approx 20(1+z)(M/M_*)^{1/3}$,  where $M_*= 3\times 10^{14}h^{-1}
M_{\odot}$.  When the correlation length is
normalized to the mean cluster separation $d_c$, it remains almost
constant: $r_c \approx$ (0.45 -- 0.5)$d_c$. For small masses of
clusters ($M<2\times 10^{14}h^{-1} M_{\odot}$) there is an indication that
$r_c$ goes slightly above the relation with the constant of
proportionality being $\approx 0.55-0.6$.

Anisotropy of density distribution in a cluster
shows no change over redshift with axial ratios
remaining constant around 1.2. In other words, clusters at present are
as elongated as they were at the epoch of their first appearance.
While the anisotropy of clusters does not change with time, the density
profile shows visible evolution: the  slope of density profile changes
from $\gamma\approx -3.5$ at $z=0.5$  to $\gamma\approx -2.5$ at the
present.  We find that the core of a cluster remains essentially the
same over time,  but the density of the  outlying regions increases
noticeably.  The virial relation $M\sim v^2$ is a good approximation,
but there is a large fraction of clusters with peculiar velocities
greater than given by this relation, and clusters with the same rms
velocities have smaller masses in the past, a factor of 2 at $z=0.5$.
\end{abstract}


\newpage
\section{Introduction}
The distribution of galaxy clusters can provide a great deal of
insight as to the structure of the universe.  By comparing with real
data such as the Abell (Abell 1958), APM (Dalton \etal 1992), or
Edinburgh-Durham (Nichol \etal 1992) cluster catalogues, the validity
of many cosmological models can be determined.
While in the past we were mainly interested in properties of clusters
at z=0, new observational results will bring more information on the
evolution of cluster properties.  Among interesting tests there are
``traditional'' tests like evolution of the correlation function and
the mass function.  It is also interesting to look for more detailed
information.  Statistics like elongation of clusters and density
profiles could shed light on the rate of evolution of fluctuations and
thus the mean density of the universe and the nature of dark matter.


Many different cosmological models have been proposed to address these
issues. Among these are the standard CDM model, CDM
with a cosmological constant $\Omega_{\Lambda}=0.5-0.8$, and the
CHDM model with 20--30\% of mass in the
form of the hot dark matter (presumably massive neutrinos).
The CHDM model (for references see Klypin \etal (1993)) was shown to
fit reasonably well
available data on cluster properties at $z=0$. For example,
Holtzman \& Primack (1993) using
the peaks formalism for gaussian density fields found that the
correlation function of clusters in the CHDM model is consistent with
the correlation function of Abell clusters. Later their
results
were confirmed by N-body simulations for Abell clusters (Jing \etal
1993, Klypin \& Rhee 1994, hereafter KR94; but see also Cen \&
Ostriker 1994) and for
APM clusters (KR94, Dalton \etal 1994). Borgani \etal
(1994, 1995) came to the same conclusion using the Zeldovich approximation.
 The CHDM model predicts (e.g., KR94, Borgani \etal
1994) the cluster mass function in
agreement with the results of White \etal (1993) and Biviano \etal
(1993) for Abell clusters. The problem of abundance of clusters with
different masses can be addressed
using other approaches. Bartlett \& Silk (1993), KR94, and Jing \&
Fang (1994) estimated the distribution function of gas temperature in
galaxy clusters using either the Press-Schechter approximation or
N-body results in conjunction with temperature--velocity dispersion
relation. The temperature distribution function was estimated by Bryan
\etal (1994) based on N-body plus hydrodynamic simulations. The
velocity dispersion function is another way to measure the abundance
of different clusters (Jing \& Fang 1994). All the tests favor the
CHDM model.
The real problem for the model is related with the amount of gas in
clusters. Estimates based on X-ray observations indicate that
the ratio of baryonic mass to the total cluster
mass is in the range $M_{\rm b}/M_{\rm tot}=0.1-0.25$ for the
assumed Hubble constant  $H_0=50{\rm km}/{\rm s}/{\rm Mpc}$ (White
\etal 1993, White \& Fabian 1995). This is
significantly larger than the upper limit $\Omega_b<0.1$ predicted from
primordial nucleosynthesis calculations (Walker \etal 1991, Krauss \&
Kernan \etal
1994). CHDM models with two or three neutrinos can slightly ease the
problem because those neutrinos are too hot and avoid the central
$\sim 1$~Mpc region of galaxy clusters. This results in decreasing of
the amount of dark matter in the central parts of clusters by at most
20--30\%, increasing the expected ratio of baryons to dark matter to
0.12--0.13.

The standard CHDM model with 30\% of mass in neutrino has severe
problems in explaining the amount of neutral gas in high redshift
($z>3$) damped Ly-$\alpha$ clouds (Klypin \etal 1995 and references
therein). More promising versions of the CHDM scenario with less mass
in the hot component (Ma \& Bertschinger 1994, Klypin \etal 1994,
Primack \etal 1995) predict significantly more high redshift
objects, making the model compatible with observational data. At the
same time, the new variants of CHDM keep the shape of spectrum and the
amplitude of fluctuations on clusters scale 10--100~Mpc almost the
same as in the old CHDM model, which still makes interesting the analysis
of clusters in the old CHDM model.

The correlation function of galaxy clusters is still an interesting
and controversial test. Recent estimates of the correlation
length for Abell clusters give $r_c\approx 20-22\Mpch$ (e.g., Peacock
\& West 1992; Postman, Huchra, \& Geller 1992), which is slightly less
than the old estimate $r_c\approx 25\Mpch$ of Bahcall \& Soneira
(1983) and Klypin \& Kopylov (1983). Results of the APM survey show
systematically lower values  $r_c\approx 13-14\Mpch$ (Dalton \etal
1992, 1994).  The difference was attributed to either
inhomogeneities in the selection of Abell clusters (e.g. Sutherland
1988) or to the difference in the richness or in the number density of
Abell and APM clusters (Bahcall \& West 1992).
 The dependence of $r_c$ on mass and number of
clusters is not well defined either in observational data or in
theoretical models.  Bahcall \&
Burgett (1986), Bahcall \& Cen (1992), and Bahcall
\& West (1992) argued that $r_c =0.4d_c$, where  $d_c$ is estimated as
n$_c^{-1/3}$, $n_c$ being the number density of clusters.
The situation with observational results is unclear. If we take
APM results (Dalton \etal 1992,  1994), EDCC results
(Nichol \etal 1992), {\it and} estimates for Abell clusters (Bahcall
\& West 1992), then all the data seems to be consistent with the
relation  $r_c =0.4d_c$. But {\it without} the Abell clusters, the APM and
EDCC results do not indicate the scaling and are consistent with no
dependence on the mean separation. Unfortunately, the range for $d_c$
is quite small in this case, so it is difficult to say if there is a
contradiction between those two results.
 From the theoretical side, the problem
is not easier.  A dependence of $r_c$ on $d_c$ or mass of clusters was
found for different cosmological models (e.g., Borgani \etal
1994,  KR94, Croft \& Efstathiou 1994). But
none of the discussed cosmological models show a perfect
$r_c \propto d_c$  relation (Croft \& Efstathiou 1994; Borgani, Coles,
\& Moscardini 1994). Moreover, results from different simulations and
different methods are somewhat contradictory.

In spite of the fact that many problems of physics of present-day
clusters  are not solved, it is quite interesting to find out what
kind of changes one expects for high-z clusters. There are some
observational results on those clusters (e.g. Gunn 1990, Luppino \&
Gioia 1995, Thimm \& Belloni (1994), Castander et al. 1994). The
situation is not clear. There are indications that some of the high-z clusters
could be just projection effects (Thimm \& Belloni 1994). Some of them
are real, but not very bright (Nichol et al. 1994, Castander et al.
1994). At the same time, it seems that there exist X-ray luminous
clusters already at $z\approx 0.5$, which can pose significant
problems for cosmological models.

This paper is organized as follows.  In section 2 the methods of
simulating the clusters and the criteria for cluster selection are
described.  The evolution of the correlation function over redshift is
covered in section 3.  Section 4 details two different methods for
measuring shapes of clusters and section 5 evaluates the evolution of virial
properties.

\section{Simulations and cluster finding algorithm}
Most of the results presented in this paper are based on
three N-body simulations.  All three were of
the standard CHDM model with $\Omega_{CDM}$=0.6, $\Omega_{\nu}$=0.3,
and h=0.5.  All simulations are scaled to the Hubble constant
H$_0$=100\begin{it}h\end{it} km s$^{-1}$ Mpc$^{-1}$.
The first two were of box length 200\h Mpc length and
are described by Klypin \& Rhee (1994).  The third simulation
was of a box with length $255\Mpch$, with a 768$^3$ mesh,
256$^3$ cold particles, and 2x256$^3$ hot particles.  This gives a
spatial resolution of 0.33\h
Mpc, with each particle having a mass of $5.43\times 10^{11}hM_{\sun}$.
The amplitude of angular fluctuations of the microwave background is
chosen in such a way that the quadrupole is normalized to $17\mu$K which is
consistent with measurements of COBE for the Harrison-Zel'dovich spectrum.
This gives the amplitude of mass fluctuations on a
8\h Mpc scale $\sigma_8 = 0.665$.  Simulations were run
from $z_0=13.2$ to the present, with positional, mass, and velocity
data saved at three epochs,
$z=0, 0.16,$ and 0.52.  The 200\h Mpc boxes have data at $z=0,
0.20,$ and 0.30.

In addition, we
use the results of one small-box high resolution simulation to address the
question of the evolution of cluster internal structure. In this case
the simulation was done for a 50\h Mpc box with a $800^3$ mesh (the
resolution is 62.5\h kpc) using 256$^3$ cold particles and 2x256$^3$
hot particles . The CHDM model with two equal mass
neutrinos accounting for $\Omega_{\rm hot}=0.20$, baryon density
$\Omega_b=0.075$, $h=0.5$, and a CMBR quadrupole of 18$\mu$K was used.

Clusters were determined as maxima of mass within a comoving radius of
1.5\h Mpc. The choice of the algorithm was motivated by the
definition of Abell clusters. We did not try to mimic the
observational procedure, specifically, projection effects. This
certainly can produce some differences by boosting ``observed'' mass
of clusters. In this paper we neglect effects of projection, which we will
address in a separate paper. Figure 1
presents an example of cluster identification. It shows two 2D
projections of cold dark matter particles and identified clusters in a
22.5\h Mpc cube from the $255\Mpch$ simulation at $z=0$. The distribution
on left panel (x-y projection) indicates that there is a quite large
supercluster, which goes almost diagonally from top right corner to
bottom left corner. Crowded distribution of a dozen of clusters around
$x =220-230~\Mpc$ nicely illustrates the difficulty of
identification of cluster in projection when supercluster is seeing
edge-on. On the right panel (y-z projection) the supercluster is
almost face-on. It is
clear that the clump of clusters is really a large filament. The plot
also shows that clusters have a strong tendency to be found in most
dense regions of the supercluster. For example, a filament at
$z=400-420~ \Mpc$, $y=395~ \Mpc$ does not have any identified
clusters with mass above $2.5\times 10^{14}h^{-1}M_{\odot}$ while its heavier
analog at $y=365~ \Mpc$ has six clusters. About half of the clusters found
in Figure 1 have visible substructure. The most massive cluster at the
top of both panel looks like a double cluster (right panel). Its mass
is $2.5\times 10^{15}h^{-1}M_{\odot}$ and its axial ratio is $a/c=1.4$ (see
section 4). There is no obvious tendency for large clusters to have
more substructure as compared with small ones. For example, the second
largest cluster on the plot with mass $1.6\times 10^{15}h^{-1}M_{\odot}$ does
look quite roundish (axial ratio
1.1-1.2). Many small clusters show significant deviation from sphericity.

Number density of
clusters in our simulations as a function of mass limit compared to
the Press-Schechter approximation and results for Abell clusters
estimated by Bahcall \& Cen (1992)
are shown in Figure 2.  If we assume that Abell  richness class
zero corresponds to a mass limit of $M> 2.5\times 10^{14}h^{-1}M_{\sun}$, then
in the $255\Mpch$ model we found 297, 121, and 13 ``Abell'' clusters
at the three redshifts 0, 0.16 and 0.52. Bahcall \& Cen's results
differ from the rest
universally by a factor of 1.6 in mass.  There are several
possibilities for this inconsistency as discussed by Klypin \& Rhee (1994).
For instance, a large portion
of the total cluster mass comes from the outer cluster regions.  Even
in the Coma
cluster uncertainties in the total mass do not accurately identify the
correct model.  The mass ranges from $0.65\times 10^{15}h^{-1}M_{\sun}$
from Bahcall \& Cen to $1.1\times 10^{15}h^{-1}M_{\sun}$ from White
\etal (1993).

\section{Evolution of correlation function of clusters}
The correlation function can be useful to describe clustering on
different scales.  We examine the correlation function over
time.  At the higher redshifts there are very few clusters that meet
our fiducial estimate $M> 2.5\times 10^{14}h^{-1}M_{\sun}$ for the mass
of a richness zero cluster. So in order to get enough clusters to do
meaningful statistics we
used half the mass limit just for this analysis.  Evolution
for the 200\h Mpc box is shown in figure 3.  Figure 4 is for the bigger
box, and it shows the same trend of increasing correlation
function at greater redshifts.  This implies that clustering {\it increased}
in the past, while the number of clusters dramatically decreased. This
result can be understood better when
the mean separation in comoving coordinates of clusters is considered.
Since the number density of clusters decreases with increasing
redshift, it is
instructive to compare the mean separation d$_c$ with the correlation
length r$_c$, defined by computing a power law $\alpha$ over the
region 2.5--25\h Mpc: $\xi=(r/r_c)^{-\alpha}$.
 The Levenburg-Marquardt
method for finding the minimum in $\chi^2$ space is used to solve
simultaneously for the correlation length $r_c$ and the slope
$\alpha$ (Press et al. 1992). In figure 5, correlation lengths are
plotted as a function of mass limit.  This displays the effects which
different mean separations have on clustering.  For all redshifts the
correlation length increases, showing increasing amplitude of the
correlation function for smaller number densities (and larger masses),
which can be seen in figures 3-4 at high redshifts.  One effect that
may be significant if real is the increase in steepness of the slope
of the r$_c$(M) relation
with redshift.  Figure 6 shows the ratio
r$_c$/d$_c$ as a function of mass limit.  This value does not equal
0.4 but remains constant near $0.46\pm 0.03$ for the $z=0$ case for
mean separations
ranging from 33.1~\h~Mpc at a mass limit of $2\times 10^{14}h^{-1}M_{\sun}$ to
49.2~\h~Mpc at a limit of $3.5\times 10^{14}h^{-1}M_{\odot}$.
Comoving correlation length $r_c$  scales with the mass limit $M$
within comoving radius $1.5\Mpch$ and the redshift $z$ approximately as
$r_c\approx 20(1+z)(M/M_*)^{1/3}$,  where $M_*= 3\times 10^{14}h^{-1}
M_{\odot}$.

\section{Shapes of clusters}
We study the evolution of cluster shapes using the 255\h Mpc
simulation at redshifts $z=$0, 0.16, and 0.52.
Two separate techniques are used.  First,
the  moment of inertia for each cluster is computed by summing
contributions for all
cold particles within the radius limit and solving for principal axes.
The ratio of the major to minor axis of the inertia tensor is used to
characterize the shapes of clusters.
Second, we examine the overall quadrupoles $Q_2$ of the clusters.
We define the multipole Q$_l$ as follows (Peebles 1980, section 46):

\begin{equation}
Q_l^2={\sum_{m=-l}^{l} a_{l}^2}, \qquad
a_l(j)={\sum_{j=1}^NY_l^m(\theta_j,\varphi_j)\over N}
\end{equation}

\noindent{}where Y$_l^m$ are spherical harmonics and the summation in
the equation  is over all dark particles $N$ within the radius limit
and over all components $m$ of the multipole.
The normalization of the multipoles $Q_l$ is defined in such a way
that the maximum value of a multipole is 1, which happens when all
particles are colinear. Typical values for a quadrupole
$Q=Q_2$ are 0.1--0.2. Both statistics are very simple to implement.
They test basically the same property, but in a slightly different way.

In Figure 7 we have split each cluster up into two sections, the inner
core ($r< 0.75\Mpch$) and the outer annulus ($0.75\Mpch < r <
1.5\Mpch$).  Poor
statistics (only 13 clusters) mar the $z=0.52$ data, but it is clear
that in all cases the outer annulus is slightly more anisotropic,
suggesting perhaps that relaxation has not been completed.
One might expect that new-born clusters at high $z$ would be less
relaxed and thus more elongated. Our results indicate that there is no
evidence for general trend with the redshift: clusters are as
elongated at high $z$ as at present.
For both epochs $z=0$ and $z=0.16$ with ample statistics the inner
core peaks at an axial ratio of 1.2 and the outer annulus at
1.3.  One effect that may show some relaxation over time is the
fact that the percentage of clusters with axial ratios $>1.4$ decreases.
At $z=0.16$ the inner and outer parts have 17.4 and 37.2\%
respectively, whereas at $z=0$ this drops to 11.7 and 22.6\%.  Figure
8 shows the quadrupole for the same clusters, again with no definite
trend.  Peaks for the three samples occur at 0.10, 0.13, and 0.17 in
increasing redshift. The distribution of quadrupoles is well
approximated by the following expression:
\begin{equation}
  {dN\over dQV}(Q)={N\over V}\left({2\over\pi}\right)^{1/2}{Q^2\over\sigma^3}
                    \exp\left(-{Q^2\over 2\sigma^2}\right),
\end{equation}
\noindent where $V$ is the total volume; parameter $\sigma$
is related to $\langle Q^2\rangle=3\sigma^2$. The rms value of the
quadrupole $\langle Q^2\rangle^{1/2}$ is 0.17, 0.20, and 0.20 for the
redshift moments 0, 0.16, and 0.52 correspondingly. The difference is
not statistically significant.

\section{Evolution of virial properties}
The virial theorem has been shown to be of use in estimating masses of
clusters.  It predicts for a cluster
$M =\alpha R v^2/G$, where $v$ is velocity dispersion, $\alpha$ is a
numerical coefficient, and $R$ is cluster radius.  In this paper we
take by definition that $R$ is the Abell
radius $R_A=1.5\Mpch$. As the result, all details of the cluster
structure are hidden in the coefficient $\alpha$. Here are some examples.
For a spherical cluster with constant density this parameter is
$\alpha=3/5$. For the King
model $\rho=\rho_0/[1+(r/r_c)^2]^{3/2}$ with truncated density at some radius
$r_{\rm max}$ the value of parameter $\alpha$ ranges from 0.6 for
$r_{\rm max}<<r_c$ to 1.21 at $r_{\rm max}=10r_c$ to 2.67 at $r_{\rm
max}=50r_c$. The general tendency is clear: larger values of $\alpha$
correspond to more centrally condensed clusters.
Figure 9 plots one-dimensional velocity dispersion of dark matter
particles against cluster mass for
all clusters in the $255\Mpch$ box with $v> 300$~km~s$^{-1}$.  No mass
limit is used, so any significant overdensity of cold particles is shown.

The majority of the clusters fall along
a line $M\propto v^2$ at the top of each graph. For $z=0$ the line
gives the value of $\alpha$ equal to 1.1, which corresponds to $r_{\rm
max}=8r_c$ if we assume the King model, which is in reasonable
agreement with our $0.33\Mpch$ resolution. The existence of the
concentration of clusters along the line is quite remarkable.
It indicates that the gravitational radius $R_g\equiv \alpha R$ is almost
the same for clusters with mass $10^{14}-10^{15}h^{-1}M_{\odot}$.
All clusters below this line have larger velocity dispersions. This might
indicate that (i) their radii are smaller than those of ``virialized''
clusters on the line, or (ii) their profiles are steeper (resulting in
larger values of $\alpha$), or (iii) significant degree of merging
activity. At first sight the latter option would seem to be the
explanation, but inspection of few cases with high velocity and low
mass shows that it might not be true. First, the most massive clusters
(mass larger than $5\times 10^{14}h^{-1}M_{\odot}$) do not deviate
that wildly from the $M\propto v^2$ line. At the same time it does not
mean that they do not show significant merging or substructure. For
example, the largest cluster in Figure 1 consists of two clusters in the
process of merging, but altogether it does not show significant
deviation from the $M\propto v^2$ line. Second, we found that clusters
with large $v$ are often those, which are close to very large nearby
clusters or are placed between massive clusters. It seems that tidal
field might significantly distort them.

It is interesting to note that
while there are many clusters with mass well below that what should be
expected for virialized clusters with a given velocity, there are no
clusters with mass {\it above} the virial relation. This is especially
remarkable if we would try to apply the usual top-hat model for cluster
formation. It nicely explains why most massive clusters are on the
$M\propto v^2$ line: their overdensity within an Abell radius is above
the 170 that is predicted by the model. But clusters with mass
$10^{14}h^{-1}M_{\odot}$ have an overdensity of only 25. If the model would
be correct those are clusters which only recently passed through
their turn-around radius and thus should have a velocity {\it
smaller} than their virial velocity. This would place them above the
$M\propto v^2$ line. In our simulations we found none of those cases. It seems
that the model works reasonably well for massive clusters, which
dominate their local environment, but fails for small clusters, which
are dominated by this environment.

In all three epochs the solid line in Figure 9 is the fit for the
$z=0$ case.  With increasing redshift the fit for each
epoch (shown as a dashed line) moves down, showing that a cluster
with the same mass had higher velocity dispersion in the past.  This
is due to a change in the overall shape of the cluster from core-dominated
at $z=0.52$ to having a larger contribution from outer regions in the
present epoch.  With more of the cluster mass  near the core in the
past, it would be easier to have an increased velocity. In order the
check this, we studied the density profile in most massive clusters.
Those with one dimensional velocity dispersions between 855 and 1000
km~s$^{-1}$ are
considered for their density profiles as shown in Figure 10.  The
average density profile has been computed at the two extreme redshifts.
For the $z=0$ nineteen clusters were used and four clusters for $z=0.52$.
Near the center the profiles are unresolved and look similar.  But
the dropoff in density happens at relatively large radii and it is
numerically resolved. It appears to be
much steeper for the clusters at large redshifts.  Two dashed
lines with power laws approaching that of the outer regions are drawn
alongside for comparison.  The dash-dot-dash line shows a fit to the profiles.
We used King approximation for the fits: $\rho/<\rho>\propto
[1+(r/r_c)^2]^{-\alpha/2}$.
For $z=0$ this fit gives $\alpha\approx 3$ and for $z=0.52$ the slope is
$ \alpha\approx 5$.

In order to check that these results are not affected by the finite
resolution of our numerical simulations, we traced the evolution of two most
massive clusters in our five times higher resolution simulation (box
size $50\Mpch$, resolution $62.5h^{-1}$kpc). Figure 11 shows the
distribution of cold dark matter inside $10\Mpch$ box around one of the
clusters at different redshifts. The cluster is slightly above the
center of the box at $z=0$. There is another (smaller) cluster few a
megaparsecs away. The fraction of particles shown for $z=0$ is twice
smaller than for the other moments. At $z=1$ the cluster is much
smaller and its center is at $x=73$ Mpc, $y=62$ Mpc (note that the axes are
not scaled with $h$). At $z=2$ the cluster is at the same position,
but it is merely a very large halo with mass around
$10^{12}M_{\odot}$. It is interesting to note how drastically the
distribution of matter changes with time. At $z=0$ most of the halos (tiny
knots of particles seeing in the Figure) are found very close to the
clusters with some indication of a filament going vertically on the
plot. At $z=0.5$ the distribution is dominated by group-size objects.
The filament is more prominent and there are some halos away from the
center. At $z=1$ the filament is the most prominent structure. Groups
are smaller, but still visible. Many halos are outside the filament.
At $z=2$ the filament has not formed yet. There are many small size
filaments and relatively large halos.

Figure 12 shows the density profile of two large
clusters ($M=4-5\times 10^{14}h^{-1}M_{\odot}$) at $z=0$ and $z=1$. Bottom
panels present the overdensity of cold dark matter particles with a distance
scale given in comoving coordinates.
Dashed curves  are analytical fits of the form
$\delta_0/(1+(r/r_c)^2)^{\alpha/2}$. For both clusters we used the
same core radii and slopes. At $z=0$ we found $r_c=0.070\Mpch$ and
$\alpha=2.30$. For $z=1$ the fits gave $r_c=0.085\Mpch$ and
$\alpha=2.70$. The core radii are still not resolved. The slope of the
density profile indicates the same trend as in our low resolution
simulations (though smaller differences): steeper profiles in the past.
In order to directly compare profiles at different redshifts we plot
the absolute density in units of mass of hydrogen atom per unit
volume in the top panels.
Radius is plotted in proper coordinates (the virial theorem
requires a proper, not a comoving radius). Now the difference in
slopes is more clear. The very central part of the cluster was better
resolved in the past: in proper coordinates the resolution
at $z=1$ is twice that at $z=0$. This explains why density is
higher at the center at $z=1$. In any case, mass within central
100~kpc radius does not change over time. What changes is the density
in peripheral parts of the clusters -- it increases by an order of
magnitude.

While the density in peripheral parts (and, thus, the total mass)
evolves quite significantly with time, the velocity dispersion is much
more stable. Figure 13 shows three-dimensional velocity profiles for
a cluster at $z=0$ and $z=1$ (full curve). Some
part of this velocity is due to streaming of material
inside the clusters. It is interesting to estimate how large that
streaming velocity is when compared with the truly chaotic component. We
estimated the chaotic velocity in the following way. We found the velocity
of each cold dark matter particle relative to the velocity of cold
matter inside a sphere of $150h^{-1}$kpc radius centered on the
particle. If there were no streaming velocities, the sphere would not
move and the rms relative velocity would be just the same as the rms
velocity of all dark matter in the cluster. In the other extreme case,
when there are large clumps moving inside clusters with relatively
small velocity of internal motion inside the clumps, we would expect
that the relative (chaotic) velocity is significantly smaller than the
total rms velocity. In reality, the situation is always somewhere in
between. The central 0.5 Mpc part seems to be quite relaxed (the two
velocities being close), while the peripheral parts show some degree
of ongoing evolution. The dashed curves in Figure 13 show the chaotic
rms velocity on a $150h^{-1}$kpc scale. At $z=1$ velocities outside the
central 0.5 Mpc region are mainly in the streaming component. At $z=0$
the chaotic velocities are much larger and dominate.

\section{Conclusions}
We have studied properties of clusters at different redshifts
predicted for the CHDM model.  Effects of the correlation function,
anisotropies, and density profiles were examined.
 We find that the correlation function
for clusters of the same mass limit was larger and steeper at high
redshifts.  The slope  increases from 1.8 at $z=0$ to 2.1 at $z=0.5$.
The comoving correlation length $r_c$  scales with the mass limit $M$
within comoving radius $1.5\Mpch$ and the redshift $z$ as
$r_c\approx 20(1+z)(M/M_*)^{1/3}$,  where $M_*= 3\times 10^{14}h^{-1}
M_{\odot}$.  When the correlation length is
normalized to the mean cluster separation $d_c$, it remains almost
constant: $r_c \approx$ 0.45 - 0.5$d_c$. For small masses of
clusters ($M<2\times 10^{14}h^{-1} M_{\odot}$) there is an indication that
$r_c$ goes slightly above the relation with the constant of
proportionality being $\approx 0.55-0.6$.

Examining isotropies we place rough limits on what is a normal or
nearly relaxed cluster.  We show that average shapes of clusters
do not change significantly over time even in the outer 1--2 Mpc parts of
clusters, where the resolution of our simulations is adequate for the
problem.  There is some evidence showing
that the small number of highly anisotropic clusters virialize
over time, but more statistics are needed here.  Five individual
clusters were examined, and those that appear normal to the eye do
statistically show to be isotropic, however selection effects on
annulus size can make an visually abnormal cluster appear normal
statistically.  We also found that the cores of clusters (inner 0.5Mpc)
remain nearly the same for redshifts $z=0-1$
while the outer regions show definite continual accretion and change
of slope of the density profile.

Finally, we can see some evidence for evolution by studying virial
properties of the clusters.  The outlying regions of clusters grow
denser at the present epoch and
have lower velocity dispersions.  Power law fits for density profiles
in $\Omega$=1 universes at $z=0$ were shown to range between 2.2 and
2.5, consistent with our results (Crone,  Evrard, \& Richstone 1994).
Our evidence
points to clusters continuing to accrete matter over time, but
retaining their basic symmetry.

%


\begin{figure}

\caption{Two 2D projections of a box with length 22.5\h Mpc from
the $255\Mpch$
sample showing cold dark matter particles.  Distances on the plot are
scaled to $h=0.5$.
Five clusters used as examples are taken from this,
ranging in mass from $2.2\times 10^{14}-1.6\times 10^{15}M_{\sun}$.
Clusters are denoted by circles surrounding
the core.  The area of the circle is proportional to the mass of the
cluster.}

\caption{A comparison of mass functions.  The cluster mass function of our $255
\Mpch$ box is shown by the solid curve.  Abell clusters estimated
by Bahcall and Cen (1992) are displayed two ways.  The
dotted curve represents their estimation, while the dashed curve
represent the mass scale increased by a factor of 1.6.  The dot-dashed
curve is the mass function predicted by the Press-Schechter
approximation with parameter $\delta_c$=1.5 and gaussian filter.}

\caption{The two point cluster-cluster correlation function $\xi$ in log (a)
and linear (b) scales for two simulations using 200\h Mpc
periodic boxes at
three different redshifts.  Vertical errors are calculated from
$\sigma=(1+\xi)/N_p^{1/2}$,
where N$_p$ is the number
of pairs counted in a given annulus.  Horizontal errors for radii less than
25\h Mpc come from $\xi$ and a power law slope computed
between 2.5 and 25\h Mpc.  Beyond 25\h Mpc horizontal
errors show the bin size which is defined as $\Delta$r/r=0.4.}

\caption{The same as Figure 2, using a different simulation involving a $255
\Mpch$ box with three times the resolution of the first sample.
As in figure 1 there is a large $\xi$ at greater redshift.  The z=0.52 curve
suffers from poor statistics, but still follows the general trend.}

\caption{Correlation length vs. the mass limit for cluster definition.
The correlation length is defined using the power law fit in the
region (2.5--25\h Mpc).
The lines for $z=0$ and $z=0.18$ represent averages of data
from two 200\h and one $255\Mpch$ simulations.  The error
bars extend from one data point to the other.  Typically twice as many
clusters in the $255\Mpch$ simulation meet
the mass limit when compared to a 200\h Mpc box.  The $z=0.3$ comes from the
200\h Mpc simulations while the $z=0.52$ comes from the $255\Mpch$
simulation.}

\end{figure}
\begin{figure}

\caption{Correlation length normalized to the mean separation between clusters.
Since the correlation function and therefore correlation length is a
function of number density, an attempt is made at normalization by
dividing the correlation length by the mean separation between
clusters.  While each sample shows some dispersion, an average of
similar redshifts shows an expected leveling off.  Errors are as in Figure 4.}

\caption{Histogram of the axial ratio between the largest and smallest
radii of the cluster.  This is determined by transforming the moment of inertia
tensor to get the principal axes and solving $I=mr^2$ for $r$.  The solid
line shows the inner volume of the cluster while the dashed line shows
the outer annulus.  The different epochs are, (a) $z=0.00$, (b)
$z=0.16$, (c) $z=0.52$.  Each epoch shows the same pattern of greater
anisotropy further from the cluster center.  The abscissa is
normalized by the volume of the box and the binsize.  One cluster is
equivalent to $1\times 10^{-6}\Mpch^{-3}$.}

\caption{Distribution of the quadrupole of the cluster anisotropy.  Epochs
and normalization are as figure 7.  The anisotropy does not appear to vary
with redshift. The dashed curves are analytical fits:
 $(N/V)(2/\pi)^{1/2}(Q^2/\sigma^3)\exp(-Q^2/2\sigma^2)$.
 Parameter $\sigma$
is related to $\langle Q^2\rangle=3\sigma^2$. The rms value of the
quadrupole $\langle Q^2\rangle^{1/2}$ is 0.17, 0.20, and 0.20 for the redshift
moments 0,
0.16, and 0.52 correspondingly. The difference is not statistically
significant. }

\caption{Mass vs. velocity at three separate epochs.  The solid line
is the fit to the z=0.0 case, while dashed lines are fits to the individual
epochs.  All clusters with velocity dispersions above 300 km s$^{-1}$
are shown.}

\caption{Average density profile for clusters with velocity dispersions
between 855 and 1000 km s$^{-1}$.  The top curve is for $z=0$ and the
bottom line for $z=0.52$.  Dash-dot-dash curves are fits to the two
profiles.  Functions are given in the text.  Dashed lines are power
laws of the given $\gamma$.}

\end{figure}
\begin{figure}

\caption{
Distribution of cold dark matter inside $10\Mpch$ box around one of the
clusters at different redshifts. The cluster is slightly above the
center of the box at $z=0$. There is another (smaller) cluster few a
megaparsecs away. The fraction of particles shown for $z=0$ is twice
smaller than for the other moments.}

\caption{Density profile of two large
clusters ($M=4-5\times 10^{14}h^{-1}M_{\odot}$) at $z=0$ and $z=1$. Bottom
panels present the overdensity of cold dark matter particles with a distance
scale given in comoving coordinates.
Dashed curves  are analytical fits
$\delta_0/(1+(r/r_c)^2)^{\alpha/2}$ with  $r_c=0.070\Mpch$ and
$\alpha=2.30$ for $z=0$. For $z=1$ the fits gave $r_c=0.085\Mpch$ and
$\alpha=2.70$. Top panels show
the absolute density in units of mass of hydrogen atom per unit
volume. Radius is plotted in proper coordinates.}

\caption{Three-dimensional rms velocity profiles for
a cluster at $z=0$ and $z=1$. Full curves are for the total velocity. Some
part of this velocity is due to streaming of material
inside the cluster.  The dashed curves show the chaotic
rms velocity on a $150h^{-1}$kpc scale. At $z=1$ velocities outside the
central 0.5 Mpc region are mainly in the streaming component. At $z=0$
the chaotic velocities are much larger and dominate.}

\end{figure}


\begin{references}

\reference Abell, G. O. 1958, \apjs, 3, 211

\reference Bahcall, N. A., \& Burgett, W.S. 1986, \apj, 300, L35

\reference Bahcall, N. A., \& Cen, R. 1992, \apjl, 398, L81

\reference Bahcall, N. A., \& Soneira, R. 1983, \apj, 270, 20

\reference Bahcall, N. A., \& West, M. 1992, \apj, 392, 419

\reference Bartlett, G.B., \& Silk, J. 1993, \apj, 407, L45

\reference Biviano, A., Girardi, M., Giuricin, G., Mardirosian, F., \&
Mezzetti, M. 1993, \apj, 411, L13

\reference Borgani, S., Coles, P., \& Moscardini, L. 1994,
\mnras, 271, 223

\reference Borgani, S., Plionis, M., Coles, P., \& Moscardini, L. 1995,
\mnras, in press

\reference Bryan, G., Klypin, A., Loken, C., Norman, M., \& Burns, J. 1994,
	\apj, 437, L5

\reference Castander, F., Ellis, R., Frenk, C., Dressler, A., \& Gunn, J.
1994, \apj, 424, L79

\reference Cen, R., \& Ostriker, J. 1994, \apj, 431, 451

\reference Croft, R.A.C., \& Efstathiou, G. 1994, \mnras, 267, 390

\reference Crone, M.M., Evrard, A.E., \& Richstone, D.O. 1994, \apj,
434, 402

\reference Dalton, G. B., Efstathiou, G., Maddox, S. J., \&
Sutherland, W. 1992, \apj, 390, L1

\reference Dalton, G. B., Croft, R.A.C., Efstathiou, G., Sutherland,
W., Maddox, S. J., \& Davis, M. 1994, \mnras, 241, L47

\reference Davis, M \& Peebles, P. J. E. 1983, \apj, 267, 465

\reference Gunn, J., 1990, in {\it Clusters of Galaxies} eds. W.
Oergerle, \etal, (Cambridge: Cambridge University Press), p.341.

\reference Hauser, M. G., \& Peebles, P. J. E. 1973, \apj, 185, 757

\reference Holtzman, J. A., \& Primack, J. R. 1993, \apj, 405, 428

\reference Jing, Y. P., Mo, H. J., Bonner, G., \& Fang, L. Z. 1993,
\apj, 411, 450

\reference Jing, Y. P., \& Fang, L. Z. 1994, \apj, 432, 438

\reference Klypin, A., Holtzman, J., Primack, J., \& Regos, E. 1993,
	Astrophys.J., 416, 1

\reference Klypin, A. A., \& Kopylov, A. I. 1983, \sovast, 9, L41

\reference Klypin, A. A., \& Rhee, G. 1994, \apj, 428, 399

\reference Klypin, A., Borgani, S., Holtzman, J., \& Primack, J. 1995,
\apj, 444, 1

\reference Krauss, L.M., \& Kernan, P.,J. 1994, \apj, 432, L79

\reference Luppino, G.A., \& Gioia, I.M. 1995, \apj, in press

\reference Ma, C., \& Bertschinger E. 1994 \apj, 434, L5

\reference Nichol, R. C., Collins, C. A., Guzzo, L., \& Lumsden, S. L.
1992, \mnras, 255, 21

\reference Peackock, J.A., \& West, M.J. 1992, \mnras, 259, 494

\reference Postman, M., Huchra, J.P., \& Geller, M.J. 1992, \apj, 384, 404

\reference Press, W. H., Teukolsky, S. A., Vettering, W. T., \& Flannery,
B. P. 1992,
Numerical Recipes in C, (Cambridge : Cambridge University Press)

\reference Primack, J., Holzman, J., Klypin, A., \& Caldwell, D. 1995,
	{\it Phys.Rev.Letters} 74, 2160

\reference Sutherland, W. J.  1988, \mnras, 234, 159

\reference Sutherland, W. J. \& Efstathiou, G. 1991, \mnras, 248, 159

\reference Thimm, G.J., \& Belloni, P. 1994, A\&A, 289,  L27

\reference Walker, T.P., Steigman, G., Kang, H., Schramm, D.M., \& Olive,
K.A. 1991, \apj, 376, 51

\reference White, S. D. M., Navarro, J. F., Evrard, A. E., \& Frenk,
C. S. 1993, Nature, 366, 429

\reference White, D.A., \& Fabian, A. 1995, \mnras, 273, 72


\end{references}
\end{document}